\documentclass[draft]{agujournal2019}

\usepackage[utf8]{inputenc}
\usepackage{xcolor}
\usepackage{amssymb}
\usepackage{amsmath}
\usepackage{float}

\usepackage{graphicx}
\usepackage{setspace}
\usepackage{bm}

\journalname{G-cubed}

\usepackage{booktabs}

\graphicspath{{./figs/}}

\begin{document}

\title{Single particle multipole expansions from Micromagnetic Tomography}
%
\authors{David Cortés-Ortuño\affil{1}, Karl Fabian\affil{2}, and Lennart V. De Groot\affil{1}}
\affiliation{1}{Paleomagnetic laboratory Fort Hoofddijk, Department of Earth Sciences, Utrecht University, Budapestlaan 17,~~3584~CD~Utrecht, The~Netherlands.}
\affiliation{2}{Norwegian University of Science and Technology (NTNU), S. P. Andersens veg 15a, 7031 Trondheim,  Norway}
\correspondingauthor{David Cortés-Ortuño}{d.i.cortes@uu.nl}
\begin{abstract}
Micromagnetic tomography aims at reconstructing 
large numbers of individual magnetizations of magnetic particles from 
combining high-resolution magnetic scanning techniques with 
micro X-ray computed tomography (microCT). 
Previous work demonstrated that dipole moments can be robustly inferred, and mathematical analysis showed that the potential field of each particle is uniquely determined. 
 Here, we describe a mathematical procedure to recover higher orders of the
 magnetic potential of the individual magnetic particles in terms of their
 spherical harmonic expansions (SHE). We test this approach on data from scanning superconducting quantum interference device microscopy and microCT of  a reference sample. For particles with high signal-to-noise ratio 
 of the magnetic scan we demonstrate that SHE up to order $n=3$ can be robustly recovered. This additional level of detail  restricts the possible internal magnetization structures of the particles and provides valuable rock magnetic information with respect to their stability and reliability  as paleomagnetic remanence carriers.  Micromagnetic tomography therefore 
 enables  a new approach for detailed rock magnetic studies on large ensembles of individual particles.
\end{abstract}

\begin{keypoints}
\item Micromagnetic Tomography   uniquely recovers higher-order multipole terms for several individual grains in a sample.
\item Higher order multipole moments are an expression of the internal domain structure of  magnetic grains.
\item Ultimately, this enables to select  individual grains for rock- and paleomagnetic studies based on domain configuration.
\end{keypoints}


\section{Introduction}

Initially, the development of micro- to nanoscale scanning magnetometers  for rock- and paleomagnetism  aimed at recovering   statistical information about the average magnetization of a sample by measuring the surface magnetic signal of the remanence carrying mineral grains \cite{Egli:2000,Weiss:2007}. One approach is to recover the total dipole moment of a larger sample volume by upward continuation of the magnetic measurements on the surface above this volume to suppress higher-order terms \cite{Fu2020}. Another approach is the spatial domain unidirectional inversion of the magnetization by means of least-squares fitting, which relies on constraints to the magnetization vector~\cite{Weiss:2007,Myre:2019}. A recently developed method of \citeA{Groot2018} improves upon the total moment measurement by aiming to recover the dipole moments of all individual magnetic particles. Although this appears to be unnecessarily complicated, this approach potentially solves almost all problems that haunt paleomagnetism from its early beginnings until today.
The most prominent amongst these problems is mineral alteration, by which the original carriers of the paleomagnetic information change in chemistry or shape and either loose their primary magnetization, or acquire a new magnetization in a different field. Another critical problem is the occurrence of   
multiple minerals as remanence carriers. If these minerals acquire their magnetization by different processes and with different efficiencies, the mixed natural remanent magnetization may not reflect the paleofield in a straightforward way.
The third common problem is a large variation in grain size or domain state of the remanence carrying magnetic mineral by which less reliable multidomain carriers may overprint and invalidate the paleomagnetic signal represented by the more reliable small grain-size fractions of pseudo-single domain or single-domain carriers.  
By individually determining the magnetic moments of all magnetized grains in a sample, it becomes possible to calculate statistical averages over specifically chosen subsets of optimal remanence carriers. For example one could remove all particles above a certain grain-size threshold from the statistical ensemble, or disregard all particles with certain  unwanted physical or chemical properties, such as all particles below a certain density threshold.
The fundamental idea that enables to recover individual particle moments independent from all other moments, is to combine micro X-ray computed tomography (microCT) and scanning magnetometry for rock magnetic measurements. This technique, called Micromagnetic Tomography, has been demonstrated by \citeA{Groot2018}. In theory, by adding microCT information, the individual magnetic potentials of topologically separated particles can be recovered from surface measurements of the normal field component on an enclosing sphere, and the corresponding inverse problem even is well-posed in the sense of Hadamard \cite{Fabian2018}. 
Here, this general result is exploited and experimentally tested by inverting not only for the magnetic dipole moments of the individual particles, but also for higher spherical harmonics, or multipole moments. One advantage of this approach is that it provides additional information about the internal magnetization structure of the particles, even though the well-known non-uniqueness of potential field inversion problems \cite{Zhdanov:2015} makes it impossible to invert for the magnetization structure itself. For example, multipole inversion may indicate that the particle magnetization is carried by a multidomain structure rather than by a single-domain structure if the magnetic scanning measurement reveals that the  quadrupole and octupole coefficients are much larger than expected for a homogenously magnetized particle. A study by \citeA{Fu2020} has recently accounted for the deviation from dipolar behavior (i.e. contribution of higher order moments) in magnetic field maps via the combination of high resolution Quantum Diamond Microscopy and upward continuation of the field data. Here, multipole moments are fully recovered from the inversions which is facilitated by knowledge of the grain positions. Multipole inversion thus may enable to intrinsically select statistical ensembles of magnetization carriers, based on their internal magnetization structure.

\section{Samples and Methods}
\subsection{Sample and tomographic characterisation}
Tomography and scanning  data in this study were acquired from
the synthetic sample described in \cite{Groot2018}. It contains natural magnetite  particles prepared and described by \citeA{Hartstra:82b}   with diameters between 5 and 35 $\mu$m. The particles were embedded in epoxy at approximately 2,800 grains per cubic millimeter. Sizes, shapes, and positions of the magnetite grains are recovered from the three-dimensional density distribution within the sample,  acquired by microCT \cite{Sakellariou:2004}. 
As described in \citeA{Groot2018}, a 1.5\,mm$\times$1.5\,mm map of the normal component of the magnetic flux density at a temperature of $T = 4$~K has been measured above the sample using scanning superconducting quantum interference device microscopy (SSM) \cite{Kirtley:1999}.
  SSM and microCT data are available from the PANGAEA data repository at https://doi.org/10.1594/PANGAEA.886724.

\subsection{Inversion method}
 The inverse modeling of the magnetic flux density is based on a forward model of the sources, where each particle center is  the center of a spherical harmonic expansion up to a maximal order $n$, where $n=1$  denotes a pure dipole potential, because magnetic potentials contain no monopole contribution. 
 In the test sample the individual particles are {\em spherically isolated} in the sense that each particle is contained in a sphere, such that no two of these spheres intersect. This is an essential requirement to ensure that the spherical harmonic potentials are in principle   uniquely defined \cite{Fabian2018}. In the following it is assumed that all sources are spherically isolated. If the individual particles do not fulfill this condition, they have to be grouped, such that the resulting groups are spherically isolated, and the recovered potentials then represent the potentials of these groups.
 
 For the forward calculation, the potentials of all sources are added and the vertical derivative of the combined potential in the scanning $x-y$-plane determines the total vertical field component $B_z(x,y,h)$, where $h$ is the scanning height of the sensor loop. The flux through the sensor is obtained by integrating this field component over the sensor area in the $x-y$-plane. By this procedure  the sensor signal in the forward model is represented by a linear combination of the spherical harmonic expansion coefficients of all particles. 
If the number of measurements is larger than the number of these expansion coefficients, the design matrix $M$  of the system becomes over-determined for the expansion coefficients, and a least-square fit of these coefficients can be obtained via the pseudoinverse of $M$.
 
A single magnetic particle inside a volume $V$ corresponds to a distribution of volume charges $\lambda(\mathbf{r})$ ($\lambda=0$ outside $V$), where $\mathbf{r}$ is the position vector within the source.  Its magnetic scalar potential at a location $\mathbf{R}$ outside the smallest sphere that contains $V$ reads

\begin{equation}
    \Phi(\mathbf{R})=\gamma_{B}\int_{V}\frac{\lambda(\mathbf{r})}{\left|\mathbf{R}-\mathbf{r}\right|}\text{d}^{3}r,
\end{equation}

\noindent where $\gamma_{B}=\mu_{0} M (4\pi)^{-1}$ and $M$ is the particle magnetization. When the observation point $\mathbf{R}$ is far away from the magnetic source, \textit{i.e.} $R \gg r$, with $R$ and $r$ the magnitudes of the position vectors, the expansion of $\left|\mathbf{R}-\mathbf{r}\right|^{-1}$ in terms of $R^{-1}$ leads to the Cartesian multipole expansion of $\Phi$ as 

\begin{align}
\gamma_{B}^{-1}\,\Phi(\mathbf{R})= & \underset{\text{dipole}}{\underbrace{(-1)\left[\int_{V}\text{d}^{3}r\,\lambda(\mathbf{r})\,r_{i}\right]\frac{\partial}{\partial R_{i}}\frac{1}{R}}}\nonumber \\
 & +\underset{\text{quadrupole}}{\underbrace{\frac{(-1)^{2}}{2!}\left[\int_{V}\text{d}^{3}r\,\lambda(\mathbf{r})\,r_{i}r_{j}\right]\frac{\partial^{2}}{\partial R_{i}\partial R_{j}}\frac{1}{R}}}\\
 & +\underset{\text{octupole}}{\underbrace{\frac{(-1)^{3}}{3!}\left[\int_{V}\text{d}^{3}r\,\lambda(\mathbf{r})\,r_{i}r_{j}r_{k}\right]\frac{\partial^{3}}{\partial R_{i}\partial R_{j}\partial R_{k}}\frac{1}{R}}}+\mathcal{O}\left(r^{4}\right) \nonumber
  \label{eq:phi-expansion}
\end{align}

\noindent where $R_i$ and $r_i$ for $ i=1,2,3$ are the Cartesian components of $\mathbf{R}$ and $\mathbf{r}$, respectively, and repeated indexes follow Einstein's summation convention. Again, the expansion starts with the dipole term because there are no magnetic monopoles. In Cartesian coordinates, all $n$th-order derivatives of $R^{-1}$ have the form
\begin{align}
    p(R_1,R_2,R_3)\,R^{-n-1},
\end{align}
where    $p$
is a  homogeneous $n$-th-order  harmonic polynomial.   The vector space spanned by  these polynomials has a basis of $2n+1$ linearly independent elements as a subspace of the space of all  $n$th-order  homogeneous polynomials.
By defining a scalar product of two polynomials $p,q$ as the average over the unit sphere by
\begin{align}
    \langle p,q\rangle ~=~ \frac1{4\pi} \int \limits_{r=1} p(\mathbf{r})\,q(\mathbf{r}) \,dS,
 \end{align}
it turns out that  the $n$-th-order derivative polynomials are not orthogonal. To correct this problem,  both the terms with derivatives and the terms with integrals are expressed as tensors. The derivative tensor is traceless for $n>1$ because $1/R$ is harmonic outside the sphere. The integral tensor defines the corresponding  multipole coefficients of the charge distribution. 

The product of these two tensors can be transformed to the basis of real spherical harmonics, which transforms the polynomials into a completely orthogonal set of basis functions. The resulting multipole expansions can be written as
\begin{equation}
    \gamma_{B}^{-1}\,\Phi(\mathbf{R})=\sum_{n=1}^{3}\sum_{\alpha=1}^{2n+1}\Theta_{\alpha}^{t(n)} \,Q^{\alpha(n)}(\mathbf{R}) +\mathcal{O}\left(r^{4}\right),
\end{equation}
\noindent where $\Theta_{\alpha}^{t(n)}$ are the components of the traceless magnetic multipole tensor of rank $n$ and $Q^{\alpha(n)}(\mathbf{R})$ are the spherical harmonic polynomials that decay as $R^{-(n+1)}$. The here applied mathematical formalism \cite{Burnham2019,Applequist2002,Stone2013}  is detailed in Appendix~A and Section~S8 of the Supplementary Material. 
The corresponding $n$-th-order multipole terms  of
the magnetic field $\mathbf{B}(\mathbf{R})=-\nabla\Phi(\mathbf{R})$   decay  proportional to $R^{-(n+2)}$. The  first (dipole), second (quadrupole) and third (octupole) order  contributions $B_{k}^{(n)}$, for $n=1,2,3$  are listed in Appendix~A.

For any maximal multipole order $n$, an approximation of the field can be constructed, where the field created by each particle at each measurement point is represented by its $n\,(n+2)$  independent multipole coefficients.
Based on these coefficients,  a $L\times\left(n\,(n+2)\,K\right)$ forward matrix is constructed that models the magnetic flux from the $K$ magnetic sources to multipole order $n$ at each of the $L$ measurement points $i$ as

\begin{equation}
B_{z|i}^{\text{(scan)}}=\sum_{j=1}^{K} B_{z|i,j}^{\text{(particle)}}=\sum_{j=1}^{K} \left[P_{z|i,j}^{\alpha(1)}\cdot\Theta_{\alpha|j}^{t(1)}+P_{z|i,j}^{\alpha(2)}\cdot\Theta_{\alpha|j}^{t(2)}+P_{z|i,j}^{\alpha(3)}\cdot\Theta_{\alpha|j}^{t(3)}\right],  
\end{equation}

\noindent with $P_{z|i,j}^{\alpha(n)}=-\partial Q_{i,j}^{\alpha(n)}/\partial z$ as the derivative of the spherical harmonic polynomials for particle $j$ at measurement point $i$ and $\Theta_{\alpha|j}^{t(n)}$ the rank-$n$ multipole tensor components of particle $j$. 
The  polynomials for the $z$-component of the magnetic field $B_{k}$  are again orthogonal. Because only  $B_{z}$ is used in the inversion, this orthogonality implies that  the SHE bases of each particle are uncorrelated, which should lead to numerically favorable  properties.

The best-fit multipole coefficients of the sources are then   computed via the pseudoinverse of this rectangular design matrix, which corresponds to the linear least-square fit of the measurement data, as long as $L> n\,(n+2)\,K$.

\section{Results}
\subsection{Tomography and magnetic scanning}

\begin{figure}
    \centering
    \includegraphics[width=0.8\textwidth]{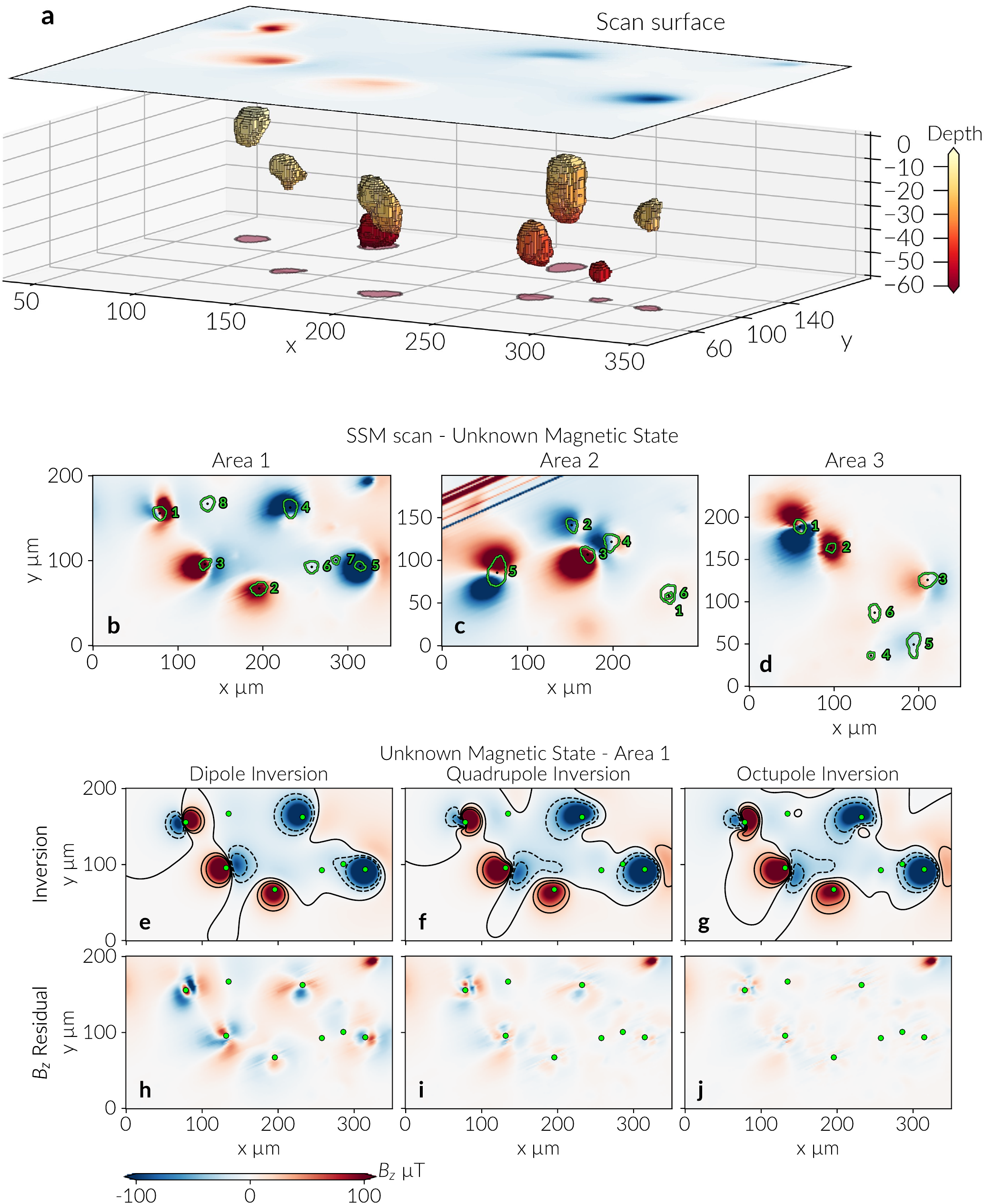}
    \caption{Multipole inversion from Micromagnetic Tomography. (a): three-dimensional view of the surface-scan region in relation to the voxel images of the magnetite grains beneath Area~1. (b-d): two-dimensional surface SSM scan data for the initial magnetic states from \citeA{Groot2018} over sample areas 1-3 with projected particle boundaries (green) from nanotomography. (e-g): stepwise  inverted $B_{z}$ fields for Area~1 of the SSM scan  using multipole expansions up to dipole (left), quadrupole (center), and octupole (right) order. Contour lines range from $-50\,\mu\text{T}$ up to $50\,\mu\text{T}$ in steps of $25\,\mu\text{T}$. (h-j):  residuals after removing the inverted $B_{z}$ fields for Area~1 from the measured data. Particle centers are indicated by green circles. }
    \label{fig:micro-grain-data}
\end{figure}

To analyze the effectiveness of the multipole expansion technique, solutions for the SSM measurements reported in \citeA{Groot2018} are obtained by solving the inversion problem in the three  areas depicted in Fig.~\ref{fig:micro-grain-data}b-d. To validate the calculations applied to real samples, test problems for the inversion of dipole signals as a function of grain position, grain depth and field scan noise, have been developed and are described in Section~S1 of the Supplementary Material. 

In \citeA{Groot2018} magnetic grains are modelled as aggregations of cubes with a constant magnetization and locations specified by microCT image analysis. This allowed to  uniquely solve for  homogeneous magnetizations in each of these spherically separated grains. Unique source assignment of the potential-field signal in this kind of system is a well-posed inverse problem \cite{Fabian2018}. The uniqueness is not restricted to the dipole moments, but extends to  the potential field of each grain, and thus to all spherical harmonic expansion coefficients. The essential constraint of the uniqueness theorem  is that the complement of the source regions must be  simply connected \cite{Fabian2018}. This excludes the possibility to reconstruct signals from source regions inside other source regions, and makes it unfeasible to obtain detailed complex magnetization structures like multi-domain structures. Yet, the multipole expansion of the potential may suffice to distinguish between a finite number of physically possible local energy minimum structures that can be modelled based on grain shape and mineralogy. The coefficients of the multipole expansion  can be found for particles for which the smallest enclosing sphere is completely below the scan surface to ensure the validity of the far-field potential description (\ref{eq:phi-expansion}) at the measurement points. These coefficients describe  point sources representing the magnetic particles. A natural choice for the origins of the multipole expansions are   the geometric centers of the particles outlined by their microCT density anomaly with respect to the surrounding matrix. Alternatively, the expansion center may be chosen as the center of the smallest sphere that contains the particle, but the  choice of the expansion center does not influence the reconstructed {\em dipole}  coefficients.

The three-dimensional image in Fig.~\ref{fig:micro-grain-data}a
visualizes the relation between the scanning surface and the position of the magnetite grains for Area~1. The plots below, Fig.~\ref{fig:micro-grain-data}b-d, 
depict the original measurement data for 
  three areas from the SSM scan together with the particle boundaries as determined by microCT.  
The   magnetic states  in the magnetic sources correspond to the initial state after preparation of the sample, CT measurement and cooling in the SQUID magnetometer,  and are considered as random magnetization states without prior information in the inversion process.
The row of images in Fig.~\ref{fig:micro-grain-data}e-g shows 
the modeled $B_{z}$ signals for the  inversions of Area~1  to stepwise increasing  multipole-expansion  orders from $n=1$ (dipole, left) to $n=3$ (octupole, right). The bottom row, Fig.~\ref{fig:micro-grain-data}h-j,  depicts the maps of the corresponding residuals. 

\begin{figure}
    \centering
    \includegraphics[width=\textwidth]{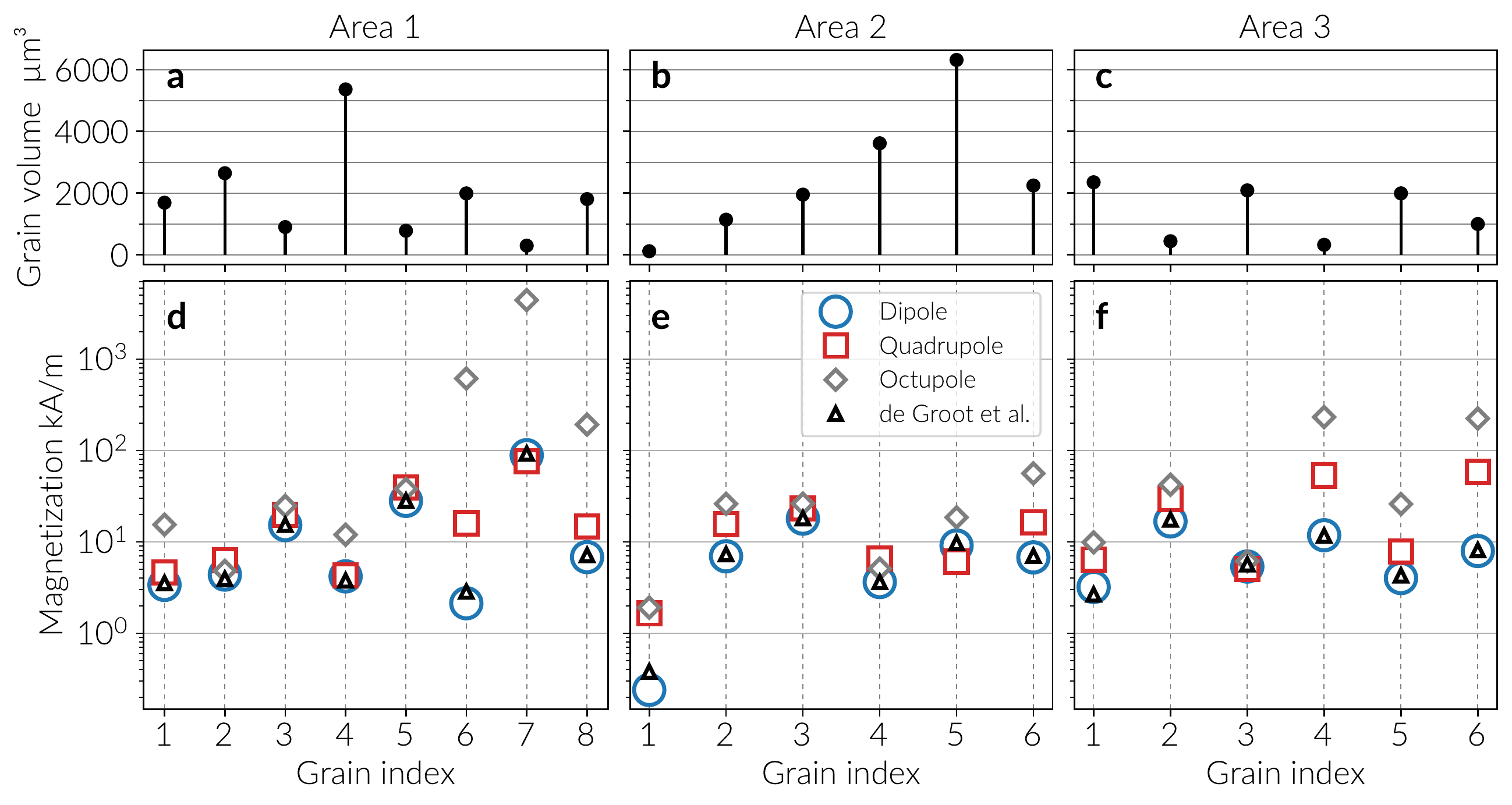}
    \caption{Grain magnetizations and volumes. Comparison of multipole inversions at different orders by means of magnetization calculations. Magnetization values are compared to those of \cite{Groot2018}. Grain volumes are reported  in the  top subplots (a)-(c).}
    \label{fig:magnetization-mult-exp}
\end{figure} 

A first test for the consistency of the inversions is to compare   the dipole moments recovered by inversions at different multipole orders. From the inverted dipole moments for the particles in  areas 1-3  the resulting average volume magnetizations are computed and  summarized in Fig.~\ref{fig:magnetization-mult-exp}a-c. The  magnetization values obtained by multipole inversion are compared to the results of \citeA{Groot2018} for the same particles. The magnetization values obtained from the lowest-order (dipole) inversion agree very well  with the homogeneous magnetization inversions of \cite{Groot2018}. In Area~1, for instance, the relative errors range from 0.17\% (grain~5) to 12\% (grain~2), with the exception of grain~6 which has a large error of 25\%. In \cite{Groot2018} each particle is represented by up to 442  cuboids to estimate its homogeneous magnetization vector. The far-field representation of grains as point-dipole sources requires substantially fewer numerical calculations and  is computationally much more effective in simulating the scan signal. 

\begin{figure}
    \centering
    \includegraphics[width=\textwidth]{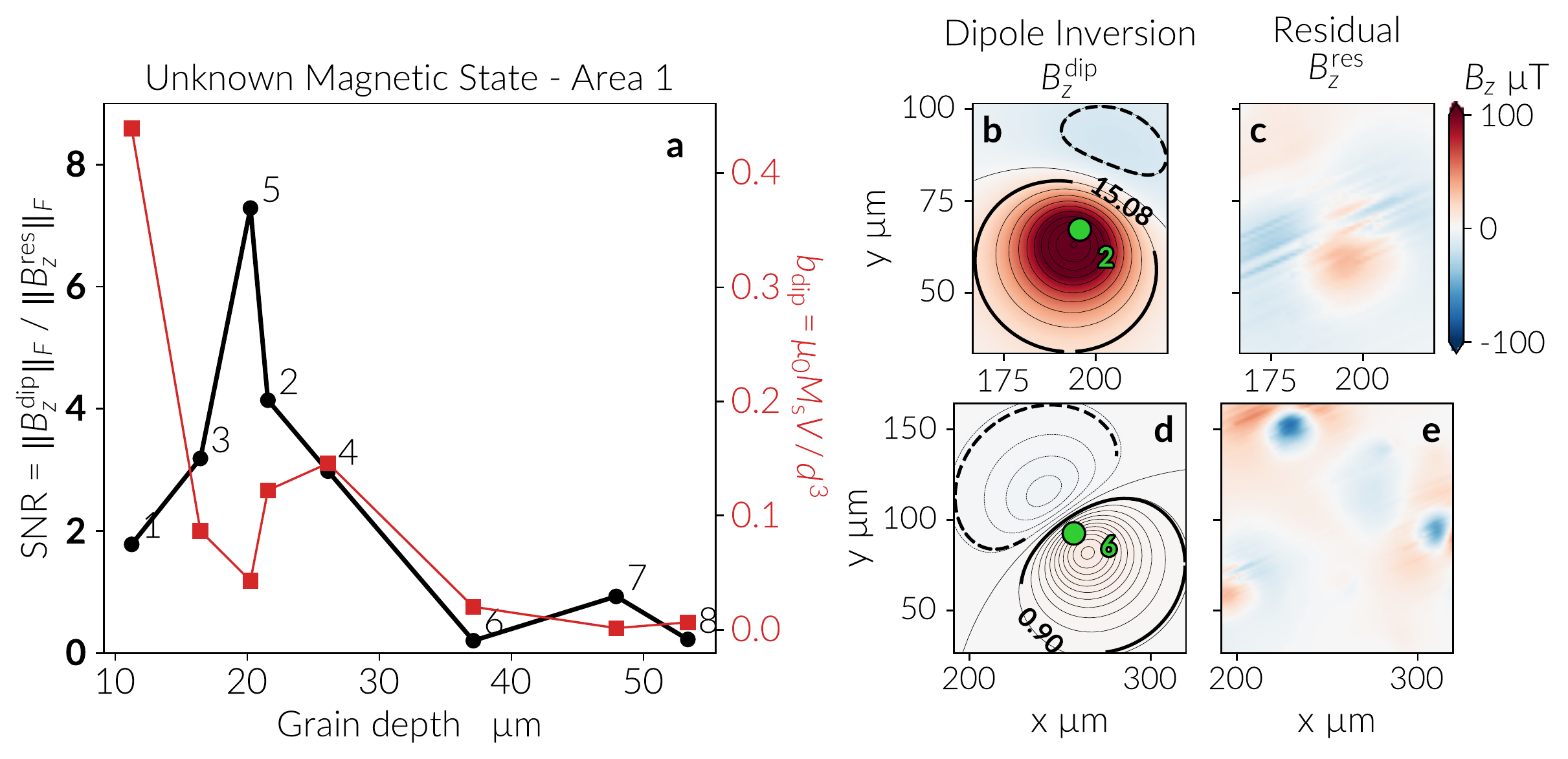}
    \caption{Signal-to-noise ratio and dipole field scaling factor as a function of grain depth. The SNR magnitudes are shown in (a) with circle markers and are computed using inversions up to the dipole order in a square area around grains that covers 90\% of the dipole signal. Both the dipole signal and residual (noise) from the inversion are shown in (b) and (c), respectively, for grain~2, and in (d) and (e) for grain~6. Contours enclosing 90\% of the positive and negative dipole signal are shown in (b) and (d). The dipole factors $b_{\text{dip}}$ are depicted in (a) with square markers and refer to the right vertical axis. Numbers at the SNR data points indicate grain indices.}
    \label{fig:SNR_depth_area1}
\end{figure}

The   residuals near the strongly expressed  grains~1-5 in Area~1 of Fig.~\ref{fig:micro-grain-data}h  are large compared to the dipole signal.  
Table~\ref{tab:inversion-results-area1} lists that for example within a radius of $30\,\mu\text{m}$ around grain~5 the largest residual value is approximately 26\% of the largest signal magnitude (see Sections~S5-S7 in the Supplementary Material for the other grains and areas). 
These high residuals demonstrate that the dipole approximation only explains a part of the measured signal, and that   higher order harmonics are required   to optimally describe the potential of the magnetic source distribution inside the particles. 
Inversion results in  Fig.~\ref{fig:micro-grain-data}f and~g belong to models with expansion up to quadrupole and octupole moments, respectively. Their residuals are shown in  Fig.~\ref{fig:micro-grain-data}i and~j. With the higher order moments the largest signal of grain~5 keeps its magnitude but the residual of the quadrupole expansion decreases to less than half the value for the dipole expansion, and remains similar for the octupole expansion (see Table~\ref{tab:inversion-results-area1}). The improvement due to  the octupole expansion is more evident in grain~1. For the quadrupole expansion, the largest residual for this grain decreases to 56\% of the maximum value of the dipolar signal, and   to 40\% for the octupole expansion. Similar tendencies are observed for grains~2-4. Although the residual does not significantly decrease with higher order terms, it is roughly an order of magnitude smaller than the  $B_{z}$ signal from the   expansion up to the octupole order.  Around grains~1 and~3 the residual after removing the octupole expansion is still 
 noticeable, with magnitudes of 72.65~$\mu$T and 65.72~$\mu$T, respectively. This   indicates that more complex multi-domain structures  are better visible in these particles because they lie closest to the scan surface (see Fig.~\ref{fig:SNR_depth_area1}). Apparently, proximity to the measurement surface   and signal strength are critical for resolving   higher order  multipole moments of the grains. This will be confirmed by analyzing the signal of grains 6 to 8.

The average volume magnetizations from the quadrupole expansion in Fig.~\ref{fig:magnetization-mult-exp}d agree well with those inferred from the dipole expansion. For grains 1-3 and 5 the relative errors are in average 38\% while grain~4 has the smallest variation with 2\% (see Table~S1 in the Supplementary Material). When octupole terms are included, deviations become significant for grains~6 and~7, and possibly grain~8. 
These grains generate the weakest signals in Fig.~\ref{fig:micro-grain-data}g and pose the greatest challenge for the numerical inversion. The residuals in  Fig.~\ref{fig:magnetization-mult-exp}i and~j for these grains are of similar size and substantially smaller than for striped residuals at other locations. This indicates that the inherent noise level inhibits an improved recovery of their higher multipole coefficients and that correlated noise leads to artificially overestimated magnetization. 

A first estimation of the quality of the inverted grain magnetizations, previous to a numerical inversion, can be deduced from the expected dipole signal. For a grain with volume $V$ at a depth $d$ from the scan surface this signal is proportional to the scaling factor (see Appendix~B for details)
\begin{equation}
    b_{\text{dip}} = \mu_{0} M_{\text{s}} \frac{V}{d^{3}}.
    \label{eq:V_R_definition}
\end{equation}
with $M_{\text{s}}$ as the saturation magnetization of magnetite. Intuitively, grains deep in the sample will have a small $b_{\text{dip}}$ which can be compensated by the grain size. The scaling factor is depicted in Fig.~\ref{fig:SNR_depth_area1}a as a function of grain depth, confirming that grains with the weakest signal, and thus with multipole inversions with large deviations, are the farthest in distance from the surface. Grain~1, on the contrary, has the strongest factor owing to its proximity to the surface, although Fig.~\ref{fig:magnetization-mult-exp} shows it suffers from deviations of the dipole moment with an octupole expansion. Similarly, grain~5 exhibits a strong scan signal, however its $b_{\text{dip}}$ magnitude is the smallest amongst the shallower grains.

Analyzing the signal-to-noise ratio (SNR) of the dipole signal, which is the strongest moment contribution, provides an extra criteria to characterize the reliability of the multipole inversion on every grain. The SNRs are shown in Fig.~\ref{fig:SNR_depth_area1}a for Area~1 of the SSM data. To calculate the SNR of a grain, a point dipole source is generated from the dipole moments of a dipole inversion (up to third order $R^{-3}$ in the expansion of the field). Then a rectangular area around the source is defined by surrounding the field contours containing 90\% of the dipole signal. This area is used as the signal matrix $B_{z}^{\text{dip}}$ and is shown in Fig.~\ref{fig:SNR_depth_area1}b and~d for grains~2 and~6, respectively. The noise matrix in this context is given by the residual of the inversion that has all dipole signals removed, and because the expansion comprises only dipolar terms the residual contains the contributions of higher order moments in addition to the measurement  uncertainties from the noise of the scan signal, which partly is spatially uncorrelated but also may be partly correlated. The ratio between the Frobenius norm $||\cdot||_{F}$ of these signal and noise matrices gives an estimation of the dipole field $B_{z}^{\text{dip}}$ energy contribution of a grain. In Fig.~\ref{fig:SNR_depth_area1}a the SNRs are plotted as a function of grain depth and it is clear that below a critical depth the SNRs decrease radically. In particular, grains~6 and~8 have SNRs smaller than one. This is in agreement with the tendency of the scaling factor $b_{\text{dip}}$. Results of SNR calculations for Area~2 and~3 (see Section S4 in Supplementary Material) also show that below a depth of approximately $30\,\mu\text{m}$ the SNRs decrease significantly. Although grain~1 of Area~1 has a strong signal, as depicted in Fig.~\ref{fig:micro-grain-data}e, its SNR is smaller than those of grains 2-5, which could be due to its proximity to the scan surface. This suggests that the far-field  approximation is probably not optimal and either an inner-field description or higher order harmonics are needed for a better description. This poorly resolved inversion also explains the difference from the large $b_{\text{dip}}$ factor that indicated a high SNR could be associated to grain~1. An additional parameter that might be affecting the inversion result is the grain size, which is shown in Fig.~\ref{fig:magnetization-mult-exp}a-c for the three SSM scan areas. It is observed that grain~7 of Area~1 has the smallest volume with a low SNR, and similarly for grain~1 of Area~2 (see Supplementary Material). Furthermore, by comparison to grain~1, grain~5 has the largest SNR with a low $b_{\text{dip}}$, suggesting that grain size can have a significant impact on the numerical inversion. Nevertheless, to statistically determine the importance of volume size in both the scan signal and thus the quality of the inversion, more experimental data is necessary.

\begin{table}[t]
    \caption{Inversion results for two grains of Area~1. Residual, inversion and Root Mean Square Error of grains~1 and ~5 within a $30\,\mu\text{m}$ radius around the grain centers. Inversions are performed using multipole expansions at  different orders.}
    \centering
    \resizebox{\columnwidth}{!}{%
        \begin{tabular}{ccccccc}
        \toprule
         & \multicolumn{3}{c}{Grain 1} & \multicolumn{3}{c}{Grain 5}\\
        \cmidrule(lr){2-4}
        \cmidrule(lr){5-7}
        Order & max(res) $\mu\text{T}$ & max($B_{z}^{\text{inv}}$) $\mu\text{T}$ & RMSE & max(res) $\mu\text{T}$ & max($B_{z}^{\text{inv}}$) $\mu\text{T}$ & RMSE\\
        \cmidrule(lr){1-1}
        \cmidrule(lr){2-4}
        \cmidrule(lr){5-7}
        Dipole & -180.08 & 308.48 & 41.53 & 74.97 & -379.75 & 19.41\\
        Quadrupole & -100.01 & 383.35 & 19.61  & -30.77 & -377.70 & 5.61\\
        Octupole & -72.65 & 410.40 & 10.84 & -37.85 & -372.24 & 4.12\\
        \bottomrule
        \end{tabular}
    }
    \label{tab:inversion-results-area1}
\end{table}

\section{Discussion}

\subsection{Inversion for the spherical harmonic expansion}

The results show that solving for higher multipole expansions of the potential of individual magnetic grains reduces the residual from the numerical inversions that recover the magnetic signal of individual grains. Considering higher multipole orders beyond the dipole approximation   improves the reconstruction of the measured signal and  refines the magnitude of the recovered grain magnetizations. 
The achievable
improvement significantly depends on the signal-quality of the  scanning measurement. The experimental data agree with the theoretical prediction  that grains exhibiting a weak signal (cf. grains 6-8 in Area~1, Fig.~\ref{fig:micro-grain-data}b) are either smaller or  located deeper in the sample, beyond the magnetometer sensitivity, which is in contrast to comparable grains with stronger signals (cf. grains 1,3,5 in Area~1, Fig.~\ref{fig:micro-grain-data}b). The multipole inversion in the here studied application becomes ineffective below a threshold depth of approximately at 30~$\mu$m beneath the sample surface. If grains  are located below each other, the signals of the deeper grains can be  shadowed even if the grains are spherically separated. This does not reflect an intrinsic non-uniqueness, but rather the fact that such constellations lead to an ill-conditioned design matrix with high noise-amplification for the deep grains. In contrast, shallow grains are generally well resolved and their inverted multipole moments are numerically stable.

\begin{figure}
    \centering
    \includegraphics[width=\textwidth]{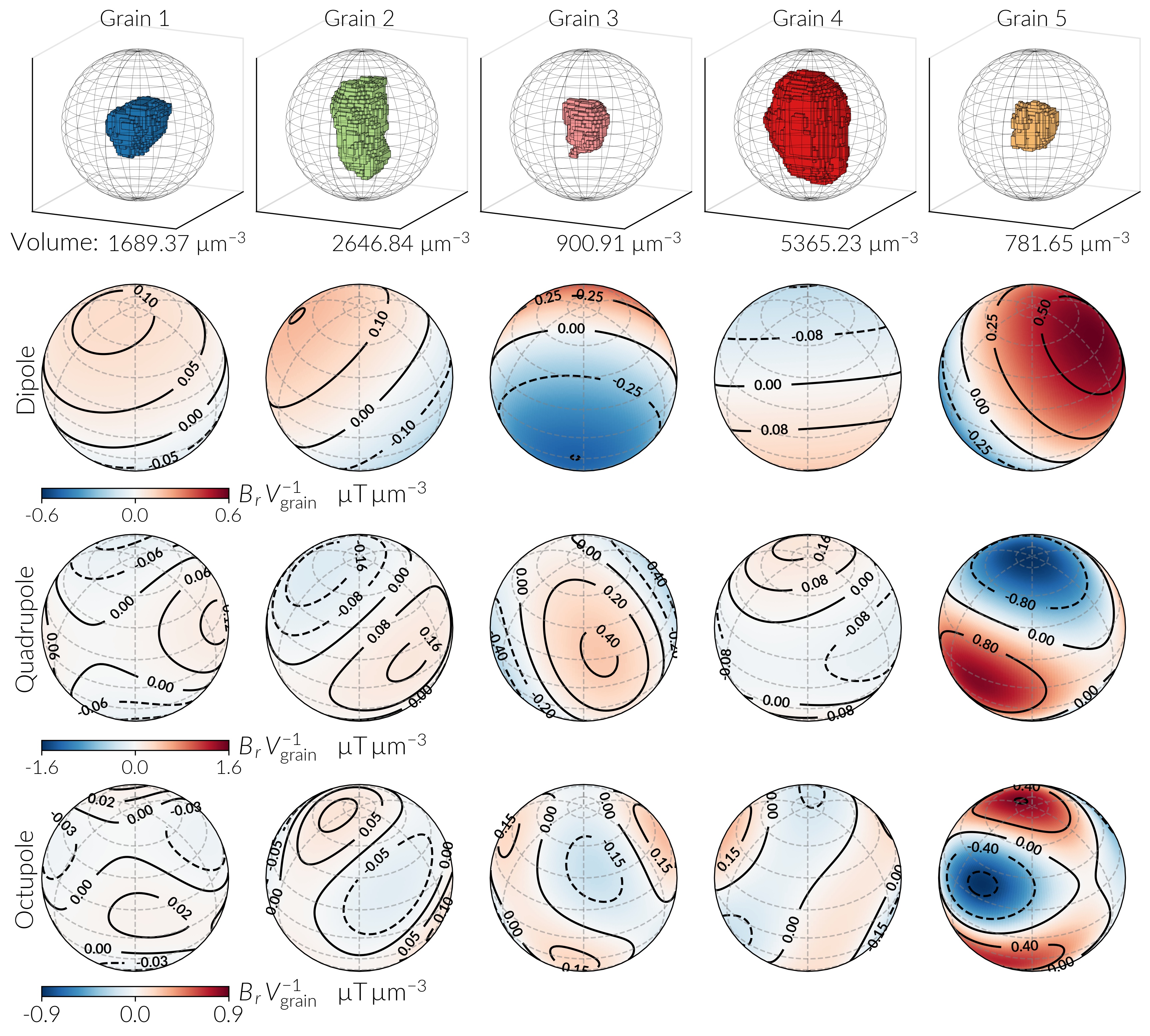}
    \caption{Grain profile and radial multipole fields of particles 1-5 in Area~1. The top row shows the geometry of the grains with bounding spheres of 20~$\mu$m, radius. Spheres are defined at the geometric center of each particle, around which the potential of the particle is represented by a SHE. The three rows below depict the radial field of each particle, reconstructed from the SSM data. The fields are orthographically projected at the surface of the bounding spheres for the first three multipole expansion terms, which are specified to the left. Field magnitudes are scaled by the particle volumes, shown below the grain profiles.}
    \label{fig:RadialMultipoleField}
\end{figure}

The inversion results were analyzed by displaying the radial field on a bounding sphere around selected individual magnetite grains. Reliable results are obtained for  grains with a strong signal that are numerically stable at inversions of increasing multipole orders. 
The radial field at the sphere surface uniquely determines the potential field  at any point outside the sphere. Thus all coefficients of the exterior spherical harmonic   expansion of the potential are determined by the radial field component, which thereby contains all outside available information about the sources within the sphere~\cite{Blakely:95}. 
Fig.~\ref{fig:RadialMultipoleField} individually depicts the corresponding radial fields for
the multipole moments of order $n=1,2$ and~3 computed from an octupole inversion
  for grains 1-5 in Area~1. 

For all grains the same radius of 20~$\mu$m was chosen for the bounding sphere. To account for the volume differences of the grains, field magnitudes are scaled by the grain volume,  specified at the top row of Fig.~\ref{fig:RadialMultipoleField}, and thus given in units of T~m$^{-3}$. This volume scaling explains why the smallest grain~5 can have the strongest field signal at all three multipole orders, with a dominant quadrupole signal. Its strong field variation at higher orders also suggests that grain~5 possesses a relatively complex magnetization structure.  A similar complexity at higher orders  is observed in grain~3 although with a weaker volume-mormalized field. Grain~4 also  exhibits a distinct  octupole character, but this is less perceptible due to its large volume, which probably allows for highly efficient flux-closure by a multi-domain magnetization structure. Also grain~1 is clearly in the multi-domain grain size region, although its higher-order fields appear less dominant than in the other grains. Its quadrupole field is of   similar order of magnitude as its dipole field, and it was observed that this grain showed a remarkable reduction of the inversion residual for $n>1$. 

\section{Conclusions}
Here we describe a new method for rock and paleomagnetism to reconstruct higher multipole moments of individual particles from surface scans of the magnetic field
in combination with micro X-ray computed tomography. The method is based on a previous mathematical proof that the corresponding linear inverse problem is well-posed. We here provide numerical and experimental evidence that the reconstruction is  technically possible, and define a set of  conditions that need to be met to robustly infer dipole, quadrupole, and octupole moments for individual particles.
We also provide analytical parameters to compare dipole and higher order moments, that allow to infer marginal domain-state information. This might be validated in future studies by means of computational micromagnetic modelling, as has been previously used to confirm complex orderings in magnetic grains from experimental imaging~\cite{Shah2018}. In this context, our study enables future rock magnetic and paleomagnetic studies to characterize and select natural mineral grains based on their domain configuration.

\acknowledgments
The data used in this study are the same data as used in \citeA{Groot2018}, and are available from the PANGAEA data repository at https://doi.org/10.1594/PANGAEA.886724.

Numerical calculations were performed using the Numpy \cite{Numpy}, SciPy \cite{Scipy}, Matplotlib \cite{Matplotlib}, Shapely \cite{Shapely} and Cartopy \cite{Cartopy} Python libraries, and symbolic calculations were obtained with Wolfram Mathematica \cite{Mathematica}.

This project has received funding from the European Research Council (ERC) under the European Union’s Horizon 2020 research and innovation programme (Grant agreement No. 851460 to LVdG).
 
\bibliography{manuscript_bib}

\begin{table}
\scriptsize
\begin{center}
\begin{tabular}{lll}
\toprule 
\addlinespace[3pt]
$p_{x}^{1(2)}=-\sqrt{\frac{3}{2}}x\left(R^{2}-5z^{2}\right)$ & $p_{y}^{1(2)}=-\sqrt{\frac{3}{2}}y\left(R^{2}-5z^{2}\right)$ & $p_{z}^{1(2)}=\sqrt{\frac{3}{2}}z\left(5z^{2}-3R^{2}\right)$\\\addlinespace[3pt]
\addlinespace[3pt]
$p_{x}^{2(2)}=-\sqrt{2}z\left(R^{2}-5x^{2}\right)$ & $p_{y}^{2(2)}=5\sqrt{2}xyz$ & $p_{z}^{2(2)}=-\sqrt{2}x\left(R^{2}-5z^{2}\right)$\\\addlinespace[3pt]
\addlinespace[3pt]
$p_{x}^{3(2)}=5\sqrt{2}xyz$ & $p_{y}^{3(2)}=-\sqrt{2}z\left(R^{2}-5y^{2}\right)$ & $p_{z}^{3(2)}=-\sqrt{2}y\left(R^{2}-5z^{2}\right)$\\\addlinespace[3pt]
\addlinespace[3pt]
$p_{x}^{4(2)}=\frac{1}{\sqrt{2}}x\left(3R^{2}-5\left(2y^{2}+z^{2}\right)\right)$ & $p_{y}^{4(2)}=\frac{1}{\sqrt{2}}y\left(7R^{2}-5\left(2y^{2}+z^{2}\right)\right)$ & $p_{z}^{4(2)}=\frac{5}{\sqrt{2}}z(x-y)(x+y)$\\\addlinespace[3pt]
\addlinespace[3pt]
$p_{x}^{5(2)}=-\sqrt{2}y\left(R^{2}-5x^{2}\right)$ & $p_{y}^{5(2)}=-\sqrt{2}x\left(R^{2}-5y^{2}\right)$ & $p_{z}^{5(2)}=5\sqrt{2}xyz$\\\addlinespace[3pt]
\bottomrule
\end{tabular}
\par\end{center}
\caption{Quadrupole field polynomials}
\label{tab:quad-field-polyn}
\end{table}

\section*{Appendix}
\subsection*{A: Magnetic field from multipole expansion}

The $k$-component of the magnetic field of an individual grain, at
a point sufficiently far away from the magnetic source, can be written
as an expansion in terms of a minimal set of real and orthogonal spherical
harmonics. This has been proved in \cite{Burnham2019}
by generalizing the results of \citeA{Applequist2002},
who defined the so called Maxwell-Cartesian spherical harmonic polynomials
that appear in the expansion of the potential using Cartesian coordinates.
Additionally, Burnham and English specify the orthogonal spherical
harmonic basis by adapting the polynomials from \cite{Stone2013}.
According to this, the field can be written as
\[
B_{k}=\sum_{n=1}^{\infty}B_{k}^{(n)}=\gamma_{B}\sum_{n=1}^{\infty}\sum_{\alpha=1}^{2n+1}\Theta_{\alpha}^{t(n)}P_{k}^{\alpha(n)}
\]
The $2n+1$ multipole polynomials $P_{k}^{\alpha(n)}=-\partial Q^{\alpha(n)}/\partial R_{k}$, which
are in the basis of spherical harmonics, can be computed by deriving
the real harmonics of \cite{Burnham2019} using Mathematica or the
open source SymPy library. By defining 
\[
p_{k}^{\alpha(n)}=R^{2n+3}P_{k}^{\alpha(n)},
\]
the three Cartesian components of the quadrupole field polynomials ($n=2$)
are summarized in Table~\ref{tab:quad-field-polyn} and the octupole field 
polynomials ($n=3$) are shown in Table~\ref{tab:oct-field-polyn}.

\begin{table}
\scriptsize
\begin{center}
\begin{tabular}{lll}
\toprule 
\addlinespace[3pt]
$p_{x}^{1(3)}=\sqrt{\frac{5}{2}}xz\left(7z^{2}-3R^{2}\right)$ & $p_{y}^{1(3)}=\sqrt{\frac{5}{2}}yz\left(7z^{2}-3R^{2}\right)$ & $p_{z}^{1(3)}=\frac{1}{\sqrt{10}}\left(3R^{4}+35z^{4}-30R^{2}z^{2}\right)$\\\addlinespace[3pt]
\addlinespace[3pt]
$\begin{aligned}p_{x}^{2(3)}= & -\frac{1}{2}\sqrt{\frac{3}{5}}\left(4R^{4}-5R^{2}\left(y^{2}+7z^{2}\right)\right.\\
 & \phantom{-\frac{1}{2}\sqrt{\frac{3}{5}}(}\left.+35z^{2}\left(y^{2}+z^{2}\right)\right)
\end{aligned}
$ & $p_{y}^{2(3)}=-\frac{1}{2}\sqrt{15}xy\left(R^{2}-7z^{2}\right)$ & $p_{z}^{2(3)}=\frac{1}{2}\sqrt{15}xz\left(7z^{2}-3R^{2}\right)$\\\addlinespace[3pt]
\addlinespace[3pt]
$p_{x}^{3(3)}=-\frac{1}{2}\sqrt{15}xy\left(R^{2}-7z^{2}\right)$ & $\begin{aligned}p_{y}^{3(3)}= & \frac{1}{2}\sqrt{\frac{3}{5}}\left(R^{4}-5R^{2}\left(y^{2}+z^{2}\right)\right.\\
 & \phantom{\frac{1}{2}\sqrt{\frac{3}{5}}(}\left.+35y^{2}z^{2}\right)
\end{aligned}
$ & $p_{z}^{3(3)}=\frac{1}{2}\sqrt{15}yz\left(7z^{2}-3R^{2}\right)$\\\addlinespace[3pt]
\addlinespace[3pt]
$p_{x}^{4(3)}=\sqrt{\frac{3}{2}}xz\left(5R^{2}-7\left(2y^{2}+z^{2}\right)\right)$ & $p_{y}^{4(3)}=\sqrt{\frac{3}{2}}yz\left(9R^{2}-7\left(2y^{2}+z^{2}\right)\right)$ & $p_{z}^{4(3)}=-\sqrt{\frac{3}{2}}\left(R^{2}-7z^{2}\right)(x^2-y^2)$\\\addlinespace[3pt]
\addlinespace[3pt]
$p_{x}^{5(3)}=-\sqrt{6}yz\left(R^{2}-7x^{2}\right)$ & $p_{y}^{5(3)}=-\sqrt{6}xz\left(R^{2}-7y^{2}\right)$ & $p_{z}^{5(3)}=-\sqrt{6}xy\left(R^{2}-7z^{2}\right)$\\\addlinespace[3pt]
\addlinespace[3pt]
$\begin{aligned}p_{x}^{6(3)}= & \frac{1}{2}\left(-3x^{2}\left(7y^{2}+z^{2}\right)+4x^{4}\right.\\
 & \phantom{\frac{1}{2}(}\left.+3y^{2}\left(y^{2}+z^{2}\right)\right)
\end{aligned}
$ & $p_{y}^{6(3)}=\frac{1}{2}xy\left(13R^{2}-7\left(4y^{2}+z^{2}\right)\right)$ & $p_{z}^{6(3)}=\frac{7}{2}xz\left(x^{2}-3y^{2}\right)$\\\addlinespace[3pt]
\addlinespace[3pt]
$p_{x}^{7(3)}=-\frac{1}{2}xy\left(28y^{2}+21z^{2}-15R^{2}\right)$ & $\begin{aligned}p_{y}^{7(3)}= & \frac{1}{2}\left(3R^{2}\left(9y^{2}+z^{2}\right)-3R^{4}\right.\\
 & \phantom{\frac{1}{2}(}\left.-7\left(3y^{2}z^{2}+4y^{4}\right)\right)
\end{aligned}
$ & $p_{z}^{7(3)}=-\frac{7}{2}yz\left(y^{2}-3x^{2}\right)$\\\addlinespace[3pt]
\bottomrule
\end{tabular}
\par\end{center}
\caption{Octupole field polynomials}
\label{tab:oct-field-polyn}
\end{table}

According to this, it is customary to obtain the radial field components as $B_{r}=B_{x}\sin\theta\cos\varphi + B_{y}\sin\theta\sin\varphi+B_{z}\cos\theta$, where the spherical angles are computed with respect to every grain center (the multipole point source).

\subsection*{B: Signal Scale}

Consider the $i$-th  particle with volume $V_i$ and saturation magnetization $M_\text{s}$ at depth $d_i$. Dependent of its magnetization structure, its average magnetization is $|\mathbf{M}_i|=\alpha_i\, M_{\text{s}}$ with a geometric factor $\alpha_i\leq 1$. Its magnetic dipole moment is therefore $\mu_i = \alpha_i\,M_\text{s}\,V_i$ and at distance $d_i$ it generates a  component $B_z$ that scales according to
\[ 
\mu_0\,\frac{\mu_i}{r_i^3} = \alpha\,\mu_0\,M_\text{s}\,\frac{V_i}{ d_i ^3}.
\]
Accordingly, the $B_z$ contribution of the $2$-nd order SHE coefficient scale proportional to 
\[ 
b_{\text{dip}}(d_i,V_i) = \mu_0\,M_\text{s}\,\frac{V_i }{ d_i^{3}}.
\]

\end{document}